\documentclass{article}
\usepackage[T1]{fontenc}
\usepackage[utf8]{inputenc}
\usepackage{textcomp}
\usepackage{eurosym}

\usepackage[textsize=footnotesize]{todonotes}

\usepackage{hyperref}
\hypersetup{
  colorlinks = true,
    linkcolor = blue,
    urlcolor  = blue,
    citecolor = blue,
    anchorcolor = blue]
}
\usepackage[backend=bibtex,style=numeric-comp,natbib=true,backref=true,url=false,doi=false]{biblatex}
\usepackage[T1]{fontenc}
\usepackage[margin=3.5cm]{geometry}
\usepackage{authblk}

\usepackage{scalerel} 
\usepackage{float}
\usepackage[labelfont=bf,font=small]{caption}
\usepackage[font=scriptsize]{subcaption}
\usepackage{mdframed}

\usepackage{graphicx}
\usepackage[printwatermark]{xwatermark}
\usepackage{xcolor}
\newwatermark[allpages,color=lightgray!50,angle=45,scale=3,xpos=0,ypos=0]{DRAFT}

\usepackage{enumitem}

\setlist[enumerate,1]{%
  label=\roman*.,
}
  
\newlist{inlinelist}{enumerate*}{1}
\setlist*[inlinelist,1]{%
  label=(\arabic*),
}

\usepackage{pdfpages}

\title{In-vehicle data recording, storage and access management in autonomous vehicles}
\author[1,2]{Viktoras Kabir Veitas%
  \thanks{Electronic address: \texttt{vveitas@gmail.com}; Corresponding author}}
\author[1]{Simon Delaere}
\affil[1]{imec-SMIT Vrije Universiteit Brussel}
\affil[2]{Center Leo Apostel - VUB}
\affil[3]{Evolution, Complexity \& Cognition Group - VUB}
\date{February 25, 2018}

\begin{document}

\maketitle
\begin{abstract}

Transport sector is in the process of being rapidly and fundamentally reshaped by autonomous and collaborative driving technologies. This reshaping promises huge economic and social benefits and brings equally huge challenges in terms of developing and deploying secure, safe vehicles and transportations systems, their smooth integration to social fabric and related adaptation of legal and regulatory framework within dynamic ecosystem. We have employed Policy Scan and Technology Strategy Design methodology \citep{veitas_policy_2018} in order to identify concrete societal expectations and problems and map them with mitigating technological availabilities in the domain of autonomous driving and smart mobility. The identified technology, having a clear place in emerging technological and regulatory landscape as well as market deployment potential is Event Data Recorder for Autonomous Driving (EDR/AD). Work on concrete junction between social problems and technology availabilities in terms of EDR/AD, presented in this article, is a continuation of the previous article on Policy Scan and Technology Strategy Design methodology \citep{veitas_policy_2018}, which are results of the same research project.

EDR/AD is an envisioned subsystem of a vehicular Controller Area Network which ensures the confidentiality, integrity and availability of data related to operation of a vehicle in order to permit recovery of exact situation following the occurrence of an event or on demand. The exact technical and regulatory requirements for the device are still in the development internationally, but it is already clear that it will be included into vehicle type-approval requirements at UNECE level. This paper presents the analysis of the context of the usage of the EDR/AD in collaborative intelligent transport systems, related security, data provenance and privacy, other regulatory and technical issues taking into account many interest groups and stakeholders involved. We present a concrete proposal for developing a EDR/AD proof of the concept prototype with clear market deployment potential and urge security researchers, vehicle manufacturers, and component suppliers to form a collaboration towards implementing important technology for making future autonomous vehicles within C-ITS more socially acceptable and legally compliant. Furthermore, EDR/AD technology, apart from its immediate use in autonomous driving and smart mobility domain has a potential to be extended to general autonomous robot and AI applications.

\end{abstract}

\section{Introduction}

One of the fastest developing areas in the domain of autonomous robotics and artificial intelligence (AI) with a potential for a large scale deployment in short or medium term is self-driving technologies and smart mobility systems based on them. The already started transition of conventional transport systems and infrastructure to collaborative intelligent transportation systems based on self-driving technologies, ubiquitous connectivity and traffic management with minimal human intervention promises the first integration of advanced AI into social life of such magnitude. Given many uncertainties inherent not only in technology developments but also in user acceptance, societal expectations and problems, changes needed to legal and regulatory systems and many more, it is instrumental for the success of the transition to ensure that the societal expectations and policies are embedded early in the technology development process.

For addressing the challenge of integrating the "application pull" determined by social needs, expectations and problems with the "technology push" originating from the accelerating development of technologies and specific solutions, we have developed the Policy Scan and Technology Strategy Design methodology. The methodology is conceived for identifying and negotiating concrete societal problems and technological solutions that mitigate them by facilitating a dialog between social world and the world of technological availabilities in immediate contexts and situations. For the in-depth presentation and discussion of the Policy Scan and Technology Strategy Design methodology see \citep{veitas_policy_2018}, which is a predecessor to the current article and is based on the results of the same research project. 

In a nutshell, the method described in \citep{veitas_policy_2018} is a design inquiry encompassing three stages which are iteratively visited many times until concrete societal problem and a technological solution mitigating it are identified and negotiated between involved parties. These stages are:
\begin{inlinelist}
  \item Content analysis (of document and interviews),
  \item Identifying / correcting directions of inquiry,
  \item Modeling the solution.
\end{inlinelist}
The goal of the methodology is to provide a systemic approach to consolidate huge informational context into socially acceptable and sustainable innovation road-map for both companies and governments. 

The current paper describes the outcome of the Policy Scan and Strategy Design process in a chosen application domain of autonomous driving and smart mobility -- which resulted in identification of \textit{in-vehicle data recording, storage and access management} technology. The identified technology is a concrete junction point between the world of current societal problems and expectations in the domain and the technology world providing availabilities to mitigate these problems.

As discussed in \citep{veitas_policy_2018}, European Union is among the most progressive governments which start to draft proposals of how to approach legal, regulatory and governance aspects related to the introduction of advanced autonomous robots and AI technologies into society (considering, among other aspects, disruption of the job market, ethical issues and the possibility that in the long term AIs will surpass human intellectual capacity). One of the ways to integrate AI to society is to pave a way for understanding independent decisions of AI through equipping all autonomous robots with a "black box" enabling recording, storing and analyzing their internal operations, logic and decisions\footnote{I.e. implementing transparency principle proposed by European Parliament \citep{european_parliament_resolution_2017}}. In case of autonomous vehicles, this requirement relates closely to the upcoming legislation requiring to upgrade the Event Data Recording devices (which are currently being installed in most cars) accommodating them to much richer and technically challenging automated driving context. Such devices would help reconstructing traffic events, understand causes of behavior of autonomous cars and autonomous traffic managers, solve legal, insurance and technical issues. Yet it also implies collecting, storing and sharing huge amounts of data at least part of which is highly privacy sensitive and regulated by international and EU data privacy and provenance laws.

In-vehicle data recording, storage and access management is a close to the market technological innovation witch addresses important legal challenges inherent in implementing intelligent transportation systems. Apart from the immediate function, this technology provides a stepping stone towards implementation of transparency principle of autonomous robots and AI in the long term. Last but not least, development and implementation of in-vehicle data recording, storage and access management systems is a business opportunity for entering dynamic and quickly growing market.

The article is structured as follows. In Section \ref{smart-mobility-market-development} we explain major developments of autonomous driving industry, relevant governance bodies and legislative domains. The section introduces relations between social and technology worlds in the autonomous driving and smart mobility domain. Section \ref{in-vehicle-data-access} addresses the legislative and technological developments, the debate of industry stakeholders around the in-vehicle data access solutions and their place in collaborative intelligent transportation systems (C-ITS). Subsection \ref{edr-ad-solution} presents in-depth analysis of a proposed concrete technological solution for solving specific societal problem having clear market deployment potential -- Event Data Recorder for Autonomous Driving (EDR-AD). Section \ref{conclusion} concludes the article and outlines further work.

\section{Autonomous driving technology and market development}\label{smart-mobility-market-development}

Technology developers, policy makers and transportation system regulators compete by proposing, experimenting and developing different visions, technical and regulatory approaches to the products and services based on self-driving technologies and their introduction into society and existing transportation infrastructures. Notwithstanding the huge developmental potential and projected social and economic benefits, self driving is only one, albeit important, technology which is disruptively changing global transportation systems. Other emerging technologies at different maturity levels are electrical power trains\footnote{e.g. Tesla -- \href{https://www.tesla.com/}{https://www.tesla.com/}}, distributed fleet management\footnote{e.g. Uber -- \href{https://www.uber.com/en-BE/}{https://www.uber.com/en-BE/}} and navigation\footnote{e.g. Waze -- \href{https://www.waze.com/}{https://www.waze.com/}}, high speed vacuum tube trains\footnote{e.g. Virgin Hyperloop One -- \href{https://hyperloop-one.com/}{https://hyperloop-one.com/}}, long distance underground tunnel networks\footnote{e.g. The Boring Company -- \href{https://www.boringcompany.com/}{https://www.boringcompany.com/}}, collaborative intelligent transportation systems\footnote{e.g. European C-ITS project -- \href{https://ec.europa.eu/transport/themes/its/c-its\_en}{https://ec.europa.eu/transport/themes/its/c-its\_en}} and drive sharing schemes\footnote{e.g. DriveNow -- \href{https://www.drive-now.com/be/en/for-businesses}{https://www.drive-now.com/be/en/for-businesses}}. Autonomous driving is an exemplar of AI technology, application of which is not exclusively related to passenger vehicles but developed with relation to all transport modalities -- trucks\footnote{e.g. Uber (former Otto) -- \href{https://www.uber.com/info/atg/truck/}{https://www.uber.com/info/atg/truck/}}, rail\footnote{\href{https://en.wikipedia.org/wiki/List\_of\_automated\_urban\_metro\_subway\_systems}{https://en.wikipedia.org/wiki/List\_of\_automated\_urban\_metro\_subway\_systems}}, marine vessels\footnote{e.g. Rolls-Royce's Advanced Autonomous Waterborne Applications Initiative -- \href{http://www.rolls-royce.com/media/press-releases/yr-2016/21-06-2016-rr-publishes-vision-of-the-future-of-remote-and-autonomous-shipping.aspx}{http://www.rolls-royce.com/media/press-releases/yr-2016/21-06-2016-rr-publishes-vision-of-the-future-of-remote-and-autonomous-shipping.aspx}} and autonomous flying vehicles\footnote{e.g. Vahana -- \href{https://vahana.aero/}{https://vahana.aero/}}. The rich and dynamic ecosystem of technological solutions direct and accelerate the development of vertically integrated technologies related to different hardware sensors, high speed data transfer networks, including new mobile networks, digital, security, software, machine learning and AI technologies -- just to name a few. These technologies and solutions dynamically interact towards creating completely new industries and transforming existing ones which cannot successfully operate without adapting legal frameworks and governance practices at almost all levels of governance. 

Many technologies, notably collaborative intelligent transportation systems cannot even be started to be developed and deployed without large scale collaborations and partnerships among infrastructure providers, policy makers and private companies. It is a considerable challenge to steer the development of highly dynamic ecosystem of technological availabilities considering the societal need for clear rules of the game in terms of legislation and standardization, which nevertheless do not stifle often unpredictable innovations.

In the area of autonomous driving and smart mobility market development, one can distinguish two trends which on the one hand compete in terms of technology availabilities and governance models, on the other -- are trying to adapt to each other:
\begin{inlinelist}
\item technologies aimed at replacing human drivers and making vehicles able to independently operate within current transport infrastructure (Subsection \ref{individual-autonomous-vehicles}) and
\item technologies aimed at interaction and collaboration of individual autonomous or semi-autonomous vehicles and by that making the whole transportation system or some of its aspects considerably more efficient and safe (Subsection \ref{c-its}).
\end{inlinelist}

\subsection{Individual autonomous vehicles}\label{individual-autonomous-vehicles}

First truly autonomous vehicles appeared in 1980s in the form of research projects by major universities and car manufacturers. In 1990s and 2000s large scale research projects in Europe and US have demonstrated self-driving cars that were able to travel thousands of kilometers in mixed traffic conditions on public roads essentially performing all driving functions. 2010s marked the accelerating commercial interest in autonomous driving technologies, with major companies (first technology companies and then established vehicle manufacturers) starting projects for developing prototypes of autonomous or semi-autonomous vehicles with clear market deployment prospects. As of 2018, all major car manufacturers as well as some technology companies have an autonomous vehicle in development and(or) already sell cars with limited self-driving capabilities (e.g. park-assist, lane change functionality, etc.). At the same time, governments, cities and traffic authorities have realized both the potential offered by the integration of self-driving technologies into transport infrastructure and public transportation systems as well as related legal, regulatory and infrastructural challenges. Currently, major regulatory initiatives for adapting traffic rules, international agreements, product and civil liability rules and more are well under way at the level of United Nations, European Union, European member states, US federal, state levels and other countries \citep{veitas_policy_2018}.

\begin{figure}[h]
  \center
  \noindent\makebox[\linewidth]{\rule{\textwidth}{0.4pt}}
  \begin{subfigure}[c]{0.49\textwidth}
    \noindent
      \includegraphics[width=1\textwidth]{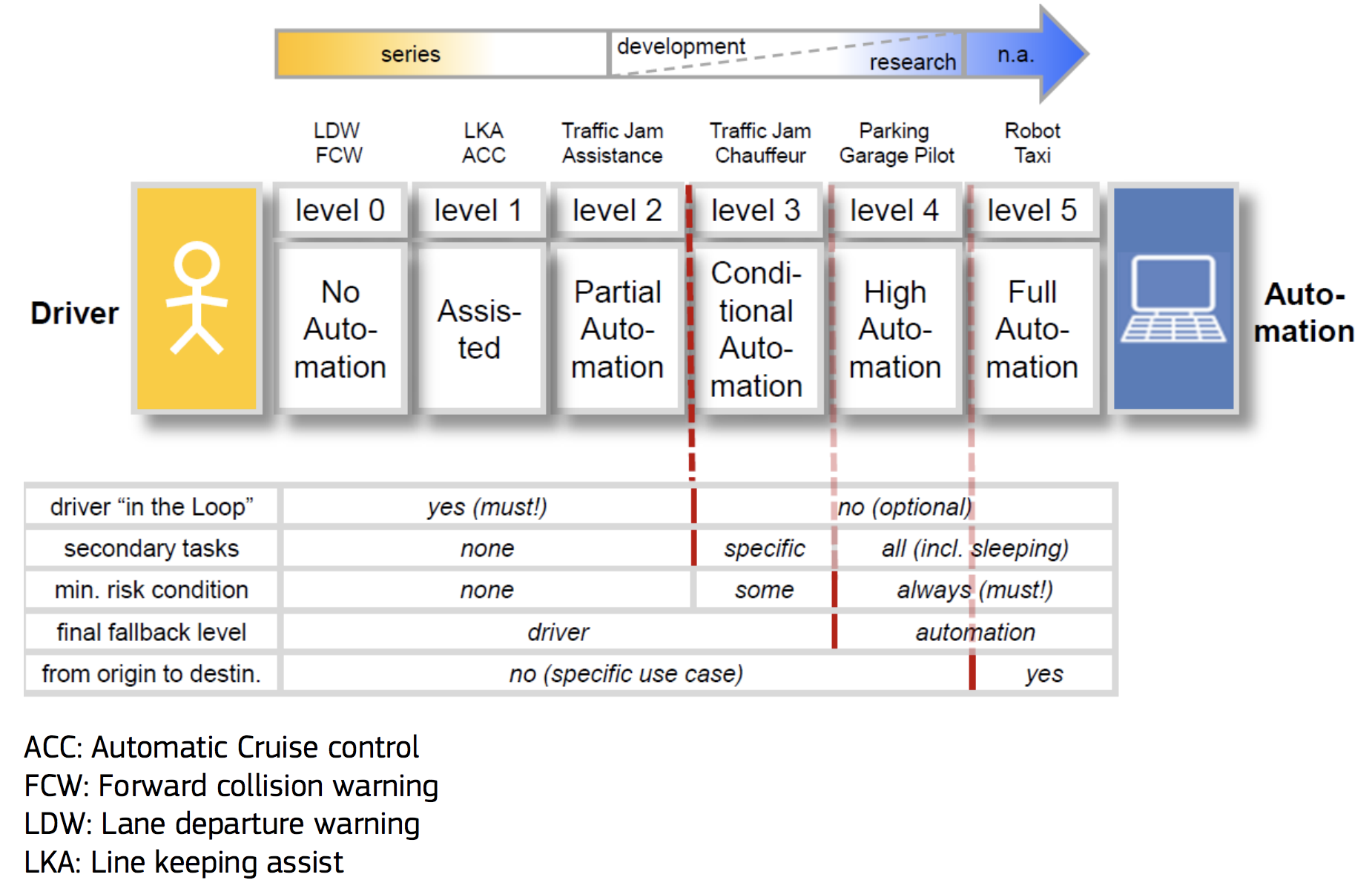}
      \caption{Different levels of automation by Society of automotive engineers (adapted from \citep[p. 42]{dg_grow_gear2030_2017})}
      \label{fig:levels_of_automation}
    \end{subfigure}%
    \begin{subfigure}[c]{0.49\textwidth}
    \includegraphics[width=1\textwidth]{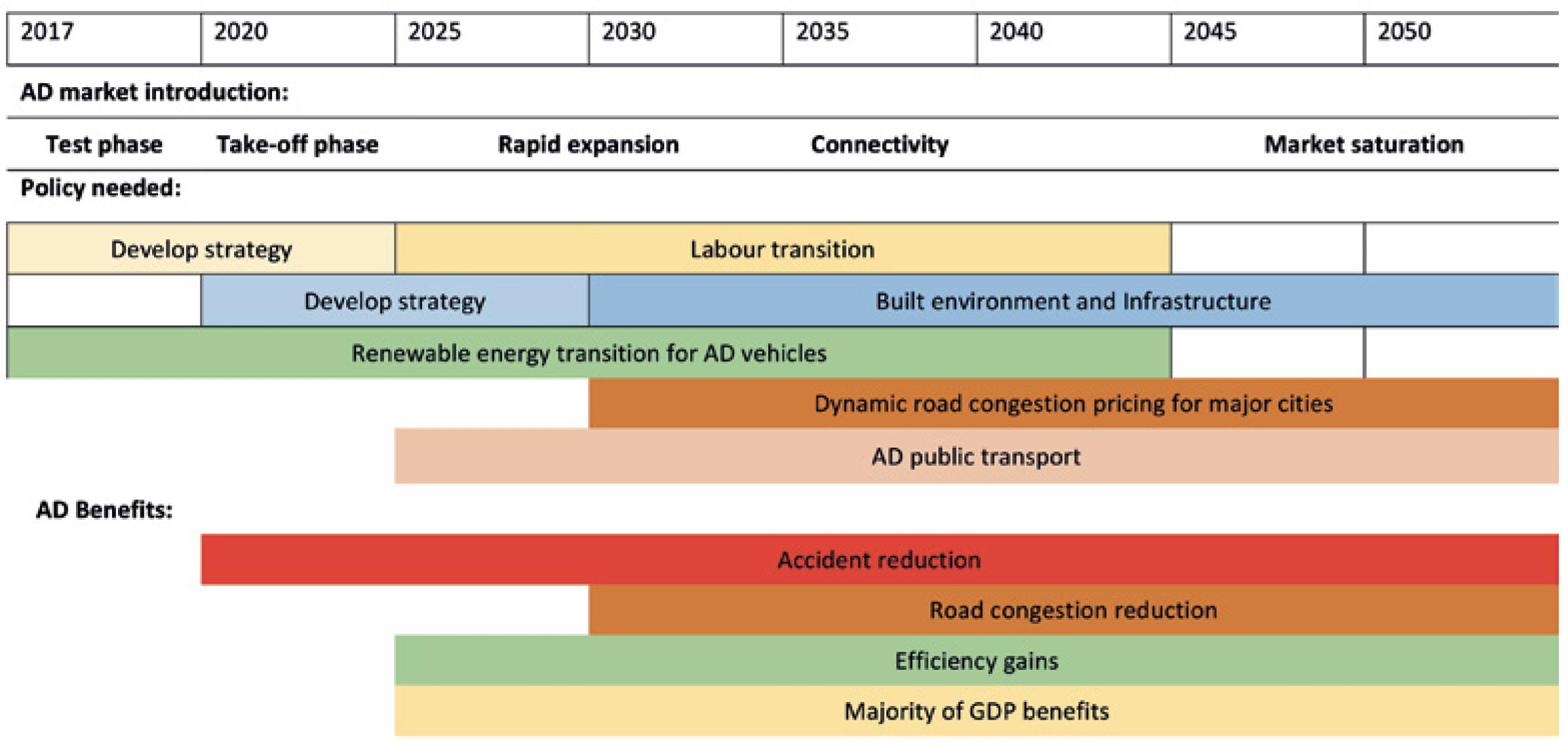}
    \caption{Timeline of AV adoption by Policy Network\footnote{\href{http://policynetwork.org/}{Policy Network, UK}} (adapted from \citep[p. 7]{ranft_freeing_2016})}
    \label{fig:timeline-of-av-adoption}
    \end{subfigure}%
\caption{SAE automation levels and estimation of the timeline of AV adoption. Note, that the time-line is very difficult to estimate given the level of unpredictability and fast pace of technological change.}
    \label{fig:av-adoption-by-sae-levels}
\end{figure}

In 2016, the International Society of Automotive Engineers proposed a taxonomy of six vehicle automation levels \citep{sae_international_j3016a:_2016} which became a \textit{de facto} standard for the industry (see Figure \ref{fig:levels_of_automation}). Level 0 refers to no automation at all, while Level 5 -- to full automation without a need for a driver in any driving conditions. Most currently produced and marketed cars feature partial automation of Level 2. It is expected that automated vehicles of SAE levels 3-4, which allow driver to perform secondary tasks for a limited number of driving scenarios, should be available in the market by 2020 (see Figure \ref{fig:av-adoption-by-sae-levels}) and fully autonomous door-to-door vehicles for any driving conditions (SAE Level 5) are expected to be available by 2030 \citep[p. 4]{article_29_opinion_2017}.

\subsection{Collaborative intelligent transportation systems}\label{c-its}

\begin{figure}[h]
  \center
  \noindent\makebox[\linewidth]{\rule{\textwidth}{0.4pt}}
  \includegraphics[width=0.9\textwidth]{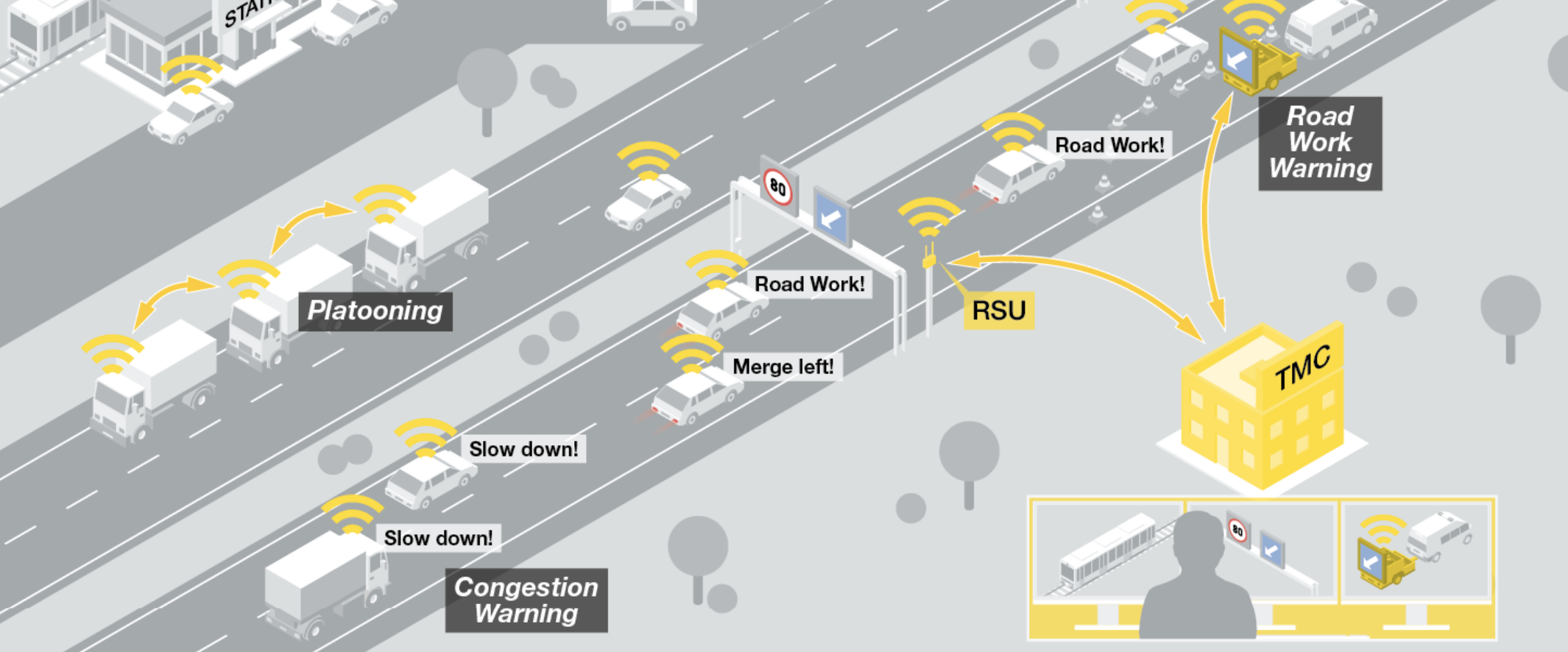}
  \caption{\textbf{C-ITS illustration}: The beacon message is known as a CAM (Cooperative Awareness Message -- shown in yellow) broadcasts position, direction of a vehicle, its speed, and other data  and forms the backbone of the analysis done by the ITS Stations in range. The result of CAM analysis and sensor data is used to detect events. If an event is detected, a DENM (Decentralized Environmental Notification Message) is sent. It specifies the event, its time and its location. Both CAM and DENM have a range of several hundred meters and can be  instantly received by all ITS Stations in range. C-ITS is a radio system, the sender has no relationship with the recipients of the messages. The messages are pseudonymised, meaning that frequently changing security certificates will veil the identity of the sender \citep[p. 12]{c-its_platform_processing_2017}.}
  \label{fig:illustration-of-c-its}
\end{figure}

Collaborative intelligent transport system (C-ITS) is a peer-to-peer solution for the exchange of data among vehicles and other road infrastructural facilities (traffic signs, traffic controllers, other transmitting/receiving base stations or smart city sensors) without the intervention of network operator (see Figure \ref{fig:illustration-of-c-its}). European Commission has adopted and is enforcing a large scale strategy for supporting the development of EU framework for more connectivity, cooperation and automation to address challenges and reap benefits on mobility in Europe. Various studies have revealed that the potential benefits of automated driving can reach up to \euro71bn in 2030 \citep[p. 42]{dg_grow_gear2030_2017}. While it is acknowledged that the potential economic and social benefits of the technological disruption are huge, they are not automatic -- the guidance and governance of the transition will have a decisive role on the ability to harness them.

The start of large scale deployment of collaborative intelligent transport system is targeted for the year 2019 in European Union. To that end, the European Commission is supporting large scale cross-border trials of automated vehicles based on smart-highway infrastructure, supports C-ITS deployment, fosters collaboration between telecoms and automotive industries, and more. It is clearly acknowledged that the first deployment of C-ITS in 2019 will cover only limited so called 'first day use cases' identified by the Commission, yet the infrastructure has to allow for gradual (but rather fast) scaling-up to advanced use cases and scenarios. 

\subsection{Relevant governance bodies}\label{governance-bodies}

The organizational structures and governance bodies which collectively shape the EU framework for connectivity, cooperation and automation of mobility sector are:
\begin{description}
  \item[C-ITS Platform.] The Platform for the Deployment of Cooperative Intelligent Transport Systems in the European Union was launched in the second half of 2014 with the intention to support the emergence of a common vision, provide and operational instrument for dialog, exchange of knowledge and cooperation on technical, legal, organizational, administrative and governance aspects \citep[p. 20]{c-its_platform_final_2017}. The platform represents all the major stakeholders of C-ITS and is comprised of nine working groups which provide outcomes in terms of technical and legal framework necessary for the deployment of C-ITS as well as needs and possibilities of higher levels of automation -- towards what is called Connected, Cooperative and Automated Mobility (CCAM) \citep[p. 8]{c-its_platform_final_2017}.

  \item[High Level Group on Automotive Industry 'GEAR 2030'] was launched by the European Commission in October 2015 in order to develop a coordinated and effective EU approach for the automotive industry at its turning point which is characterized by the need to embrace the upcoming digital revolution, automated and connected driving, environmental challenges, societal changes and globalization \citep[p. 5]{dg_grow_gear2030_2017}.
  
  \item[United Nations Economic Commission for Europe (UNECE)] houses the Global Forum for Road Traffic Safety (Working Party 1)\footnote{\href{http://www.unece.org/trans/main/welcwp1.html}{http://www.unece.org/trans/main/welcwp1.html}}, World Forum for Harmonization of Vehicle Regulations (Working Party 29)\footnote{\href{https://www.unece.org/trans/main/wp29/meeting\_docs\_wp29.html}{https://www.unece.org/trans/main/wp29/meeting\_docs\_wp29.html}} and also regulates uniform provisions concerning the approval of vehicles with regard to steering equipment (UN Regulation 79)\footnote{\href{https://wiki.unece.org/pages/viewpage.action?pageId=25265606}{https://wiki.unece.org/pages/viewpage.action?pageId=25265606}}. The type approval and road worthiness rules applicable in EU are developed by these bodies.

  Specifically, Working Party 29 (WP.29)\footnote{\href{https://www.unece.org/trans/main/wp29/introduction.html}{https://www.unece.org/trans/main/wp29/introduction.html}} is a worldwide regulatory forum which provides the legal framework allowing member countries to establish regulatory instruments concerning motor vehicles and equipment. The framework developed by WP.29 allows the market introduction of innovative vehicle technologies, while continuously improving global vehicle safety via defining performance test and administrative requirements and procedures, type approvals, the conformity of production, periodic technical inspections and more.
  
  \item[Article 29 Data Protection Working Party]\footnote{\href{http://ec.europa.eu/newsroom/article29/item-detail.cfm?item_id=605262}{http://ec.europa.eu/newsroom/article29/item-detail.cfm?item\_id=605262}} deals with data protection issues for European Commission and Council and is comprised out of representatives of data protection supervisory authorities of each member country, representatives of relevant EU institutions and European Commission.
  
  \item[European Standards Organization (ETSI)] - a regional standards body dealing with telecommunications, broadcasting and other electronic communication networks and services. The EU explicitly recognizes the standartization procedures of ETSI as technically thorough, inclusive and permits to draft European Standards, EN standards and lends them their legitimacy \citep[p. 7]{c-its_platform_processing_2017}.

  \item[ISO Technical Committee (TC) 204] "is responsible for the overall system aspects and infrastructure aspects of intelligent transport systems (ITS), as well as the coordination of the overall ISO work programme in this field including the schedule for standards development, taking into account the work of existing international standardization bodies"\footnote{\href{https://www.iso.org/committee/54706.html}{https://www.iso.org/committee/54706.html}};

  \item[European Committee for Standardization (CEN)]\footnote{\href{https://www.cen.eu/Pages/default.aspx}{https://www.cen.eu/Pages/default.aspx}} is officially recognized by EU as European standards (ENs) body and is developing ENs for building European (Digital) Single Market. CEN Technical Committee 278\footnote{\href{https://standards.cen.eu/dyn/www/f?p=204:7:0::::FSP_ORG_ID:6259&cs=1EA16FFFE1883E02CD366E9E7EADFA6F7}{https://standards.cen.eu/dyn/www/f?p=204:7:0::::FSP\_ORG\_ID:6259\&cs=1EA16FFFE1883E02CD366E9E7EADFA6F7}} is developing all standards related to Intelligent Transport Systems in Europe;

  \item[SAE International] (initially established as Society of Automotive Engineers)\footnote{\href{https://www.sae.org/}{https://www.sae.org/}} is professional organization and standards developing body based in US with principal emphasis on transport industries.

\end{description}

Discussions of industry stakeholders and policy makers within these bodies shape the collective decisions and eventual legislation of collaborative intelligent systems and automated driving in Europe and internationally. By considering all aspects of the industry transition, the emerging regulatory framework touches and in part guides specific technological solutions. Implications of the framework to the in-vehicle data recording, storage and access management technology are discussed in Section \ref{edr-ad-solution}.

\subsection{Relevant legislative domains}\label{relevant-legislative-domains}

The five large legislative domains which are related to the future smart mobility solutions and connected and autonomous driving technologies as they will be deployed in European market are: 
\begin{description}
  \item[Product liability framework] is implemented by Product Liability Directive in EU. The directive lays down common rules for the liability without fault of the producer for damage caused by a defective product. The aim of the directive is to facilitate the free movement of goods while ensuring an equal level of protection for consumers. Autonomous robots in general and self-driving cars in particular, due to their ability to learn and actuate new behaviors, make it difficult to trace liability to owner, user or manufacturer in case of an accident or inappropriate behavior. Although the proposals how to deal with this challenge vary from suggestions for establishing electronic personality for autonomous robots \citep[p. 16]{european_parliament_resolution_2017} to applying the current framework without change, the current opinion of experts is that product liability framework need not change in the nearest future\footnote{See \href{https://ec.europa.eu/digital-single-market/en/news/workshop-liability-area-autonomous-systems-and-advanced-robots-and-internet-things-systems}{proceedings} of the Workshop on liability in the area of autonomous systems and advanced robots and Internet of Things systems (at European Parliament)}.
  \item[Type approval framework] is implemented by Type Approval Rules in EU which are developed by UNECE. Type approval describes the process applied by national authorities to certify that a model of a vehicle meets all EU safety, environmental and conformity of production requirements before authorizing it to be placed on the EU market. Type approval rules as well as certification procedures will definitely need to be adapted for the autonomous cars and autonomous robots as well as their infrastructure -- a process that is still well under way.
  \item[Data Privacy and Provenance Policy] is a part of the initiative of building a European data economy\footnote{\href{https://ec.europa.eu/digital-single-market/en/policies/building-european-data-economy}{https://ec.europa.eu/digital-single-market/en/policies/building-european-data-economy}} which is in turn a part of European Digital Market Strategy and is implemented by two legislative documents: General Data Protection Regulation (GDPR)\footnote{\href{http://www.eugdpr.org/}{http://www.eugdpr.org/}} and Regulation on Privacy and Electronic Communications (e-Privacy Regulation)\footnote{\href{http://www.europarl.europa.eu/legislative-train/theme-connected-digital-single-market/file-e-privacy-reform}{http://www.europarl.europa.eu/legislative-train/theme-connected-digital-single-market/file-e-privacy-reform}}, both of which are set to be put into force on 25 May 2018. GDPR provides legal protection of personal data of natural persons and applies to all companies or organizations collecting and/or processing personal data of data subjects residing in EU. Apart from the serious penalties which will be imposed on companies failing to comply with the regulation, it establishes rules for collecting \textit{explicit consent} of data subjects to process their data, \textit{breach notifications}, \textit{right to access} (information and procedures about personal data being processed) by data subjects, \textit{right to be forgotten} (i.e. erased from the personal data and cease its further dissemination), \textit{data portability} (i.e. ability to transparently change company or organization which processes the personal data) and \textit{privacy by design}. E-Privacy Regulation augments GDPR by dealing with protecting private data in electronic communications (including machine-generated data and machine-to-machine communications) in terms of enhancing security and confidentiality, defining clearer rules of tracking technologies, addressing fragmentation of legislation in the domain as well as consistent enforcement with agreement to GDPR.
  \item[Free flow of data framework] is also a part of European data economy initiative and is implemented by the Regulation of the European Parliament and the Council on a framework for the free flow of non-personal data in the European Union (free flow of data regulation)\footnote{\href{http://www.europarl.europa.eu/legislative-train/theme-connected-digital-single-market/file-free-flow-of-data}{http://www.europarl.europa.eu/legislative-train/theme-connected-digital-single-market/file-free-flow-of-data}}. The regulation defines rules for storage and processing of non-personal data across the European Union including machine generated data and together with Data Privacy and Provenance Policy cover all production, storage and communication of all types of data. An important note is in place here: despite the impression of structure that the adopted and proposed legislation for data governance provides, there is great uncertainty about how exactly it will play out when exposed to the real world situations and cases. For example, whether specific data is 'personal', 'sensitive' or 'non-personal' (and therefore which regulation is applicable) is very much context dependent -- especially considering quickly developing machine learning and so called Big Data analysis techniques which allow to identify person from mining seemingly 'non-personal' data. Therefore the appropriate way to approach the regulatory structure of European data economy is treat it as guidelines and framework of cooperation between involved stakeholders (policy makers, businesses and data subjects) rather than strictly established and clear rules to be followed. 
  \item[C-ITS Certificate and Security Policy] aims to define a trust model in Europe to ensure secure and interoperable exchange of C-ITS messages on EU-wide level. The model is being developed by the Working Group Security of C-ITS Platform which has released the first agreed upon by all involved stakeholders document in June 2017\footnote{\href{https://ec.europa.eu/transport/sites/transport/files/c-its_certificate_policy_release_1.pdf}{https://ec.europa.eu/transport/sites/transport/files/c-its\_certificate\_policy\_release\_1.pdf}}. The document defines the European C-ITS trust model based on Public Key Infrastructure (PKI) as well as details on the roles and processes of how security certificates are issued including schemes for ensuring pseudonymity via authorization tickets. Importantly, data access and provenance solutions will have to be integrated with the PKI trust model being developed.
\end{description}

Despite the breath and scope of these legislative domains, they directly influence the in-vehicle data recording, storage and access management technology which we discuss further.

\section{In-vehicle data access}\label{in-vehicle-data-access}

Deployment of autonomous and collaborative driving technologies will generate unprecedented amounts of data flows, management of which will strongly influence operational efficiency, security and overall socio-economic benefits of the system as well as smoothness of the transition. The higher the automation of a vehicle, the more important are data flows among internal components to their function and operation -- presenting considerable technical, regulatory, security and safety challenges. Collaborative intelligent transport (C-ITS) system, where data is being exchanged between vehicles and can influence their decisions, adds yet another layer of complexity to the already challenging problematics. 

Access to in-vehicle data has a specific position within the C-ITS platform, since its scope is beyond purely C-ITS and includes existing and possible future in-vehicle applications or services. Certain parts of data will have to be stored for prolonged periods of time and/or made available not only to competent authorities, but also for other parties of C-ITS system (car manufacturers, users, operators, service providers, etc.), especially beyond first deployment of the system. Besides security, data privacy and provenance issues will become relevant as long as in-vehicle storage will be accessible and potentially copy-able for third parties. Personal data  should  be  appropriately secured,  including protection  against unauthorized  processing and  against disclosure, accidental  loss, destruction or  damage, using appropriate technical and organizational  measures (‘integrity  and confidentiality’) \citep[p. 18]{c-its_platform_processing_2017}.  

In this section we discuss these challenges in detail and introduce and explain Event Data Recorder for Automated Driving (EDR-AD) -- the proposed technical solution for mitigating concrete (albeit of course not all) challenges of integrating automated driving technologies into society. See the predecessor of the current article \citep{veitas_policy_2018} for the explanation of Policy Scan and Technology Strategy design process that led to targeting in-vehicle data storage domain and identification of the EDR/AD solution as the best candidate for short/medium term market deployment.

\subsection{In-vehicle data recording, storage and access management requirements}\label{in-vehicle-data-storage-requirements}

Security and safety of autonomous vehicles and collaborative intelligent transportation systems is a strict overarching requirement for their market deployment and is a formidable technological and governance challenge. Real-time in-vehicle data recording, storage and access management is instrumental for reconstructing traffic events and understanding what caused them, also for continuous improvement of technology. Five guiding principles should be applied when granting access to the in-vehicle data resources\citep{c-its_platform_working_group_6_access_2015}:
\begin{description}
  \item[Data provision conditions: consent.] The data subject (owner of the vehicle and/or through the use of the vehicle or nomadic devices) decides if data can be provided and to whom, including the concrete purpose for the use of the data (and hence for the identified service). There is always an opt-out option for end customers and data subjects. This is without prejudice to requirements of regulatory applications.
  \item[Fair and undistorted competition.]Subject to prior consent of the data subject, all service providers should be in an equal, fair, reasonable and non-discriminatory position to offer services to the data subject.
  \item[Data privacy and data protection.] There is a need for the data subject to have its vehicle and movement data protected for privacy reasons, and in the case of companies, for competition and/or security reasons.
  \item[Tamper-proof access and liability.] There is a need for the data subject to have its vehicle and movement data protected for privacy reasons, and in the case of companies, for competition and/or security reasons.
  \item[Data economy.] Given all above, standardized access favors interoperability between different applications and facilitates the common use of same vehicle data and resources \citep[p. 7]{c-its_platform_working_group_6_access_2015}.
\end{description}

The exact in-vehicle data access and sharing solution that will be deployed within C-ITS will be determined not only by technological availability but also as a result of negotiation between multiple interest groups within the industry -- vehicle manufacturers, component suppliers, repair and maintenance, testing and certification, application service, IT service providers, road authorities, vehicle rental and fleet operators and policy makers.

There are three main technical solutions currently negotiated among industry participants and interest groups: 
\begin{description}
  \item[On-board application platform:] internal in-vehicle operating system that supports installing third-party applications which access to the data of the vehicle;
  \item[In-vehicle interface] which provides an interface for applications (internal or external) to access in-vehicle data;
  \item[Data server] where all data is uploaded to the server outside a car which manages  data access to third parties and applications. Furthermore, three data server based solutions are distinguished: extended vehicle, shared/neutral and B2B marketplace.
\end{description}

\begin{figure}[H]
  \center
  \noindent\makebox[\linewidth]{\rule{\textwidth}{0.4pt}}
  \begin{subfigure}[c]{0.48\textwidth}
    \noindent
      \includegraphics[width=1\textwidth]{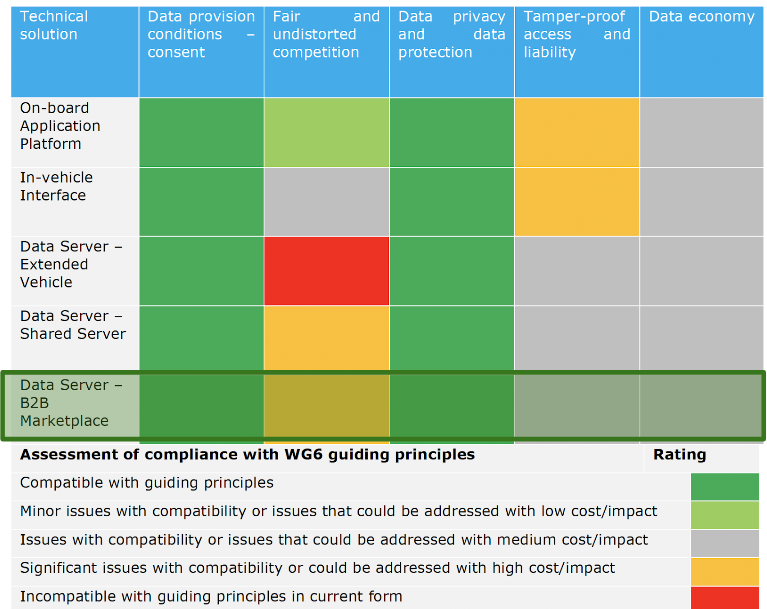}
      \caption{Compliance of different solutions to the five guiding principles.}
      \label{fig:guiding-principles}
    \end{subfigure}%
    \quad
    \begin{subfigure}[c]{0.48\textwidth}
    \includegraphics[width=1\textwidth]{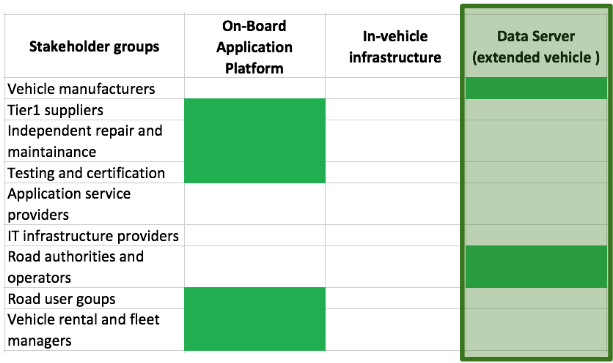}
    \caption{Preferences of industry interest groups towards the candidate in-vehicle data access solutions.}
    \label{fig:stakeholder-preferences}
    \end{subfigure}%
\caption{Analysis of proposed in-vehicle data access architecture for implementation in C-ITS scenarios. Green rectangles indicate the preferred solution by each interest group, shaded area in green -- the solution with strongest support. Adapted from \citep{mccarthy_access_2017}.}
    \label{fig:solution-analysis}
\end{figure}

\citet{mccarthy_access_2017} provides the analysis of these candidate solutions in light of guiding principles required by policy makers and preferences of industry interest groups (see Figure \ref{fig:solution-analysis}) and concludes that the data server concept has the largest potential to be implemented in C-ITS. This conclusion was made on the grounds of consideration that if policy makers will not intervene into the market, vehicle manufacturers will implement their preferred solution, which is data server. Considering 
\begin{inlinelist}
\item largely "wait and see" policy of the European Union towards new technologies so that not to stifle innovation and 
\item the short time-frame for first C-ITS implementation in Europe,  
\end{inlinelist}
market intervention is very unlikely, therefore data server architecture is the most probable.

\begin{figure}[h]
  \center
  \noindent\makebox[\linewidth]{\rule{\textwidth}{0.4pt}}
  \includegraphics[width=0.7\textwidth]{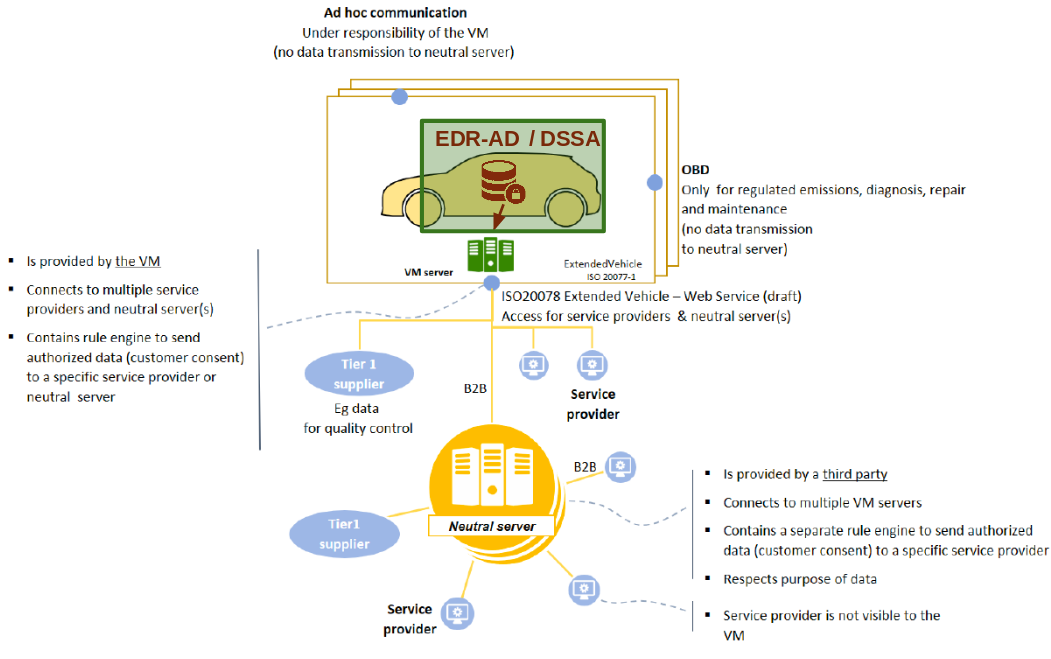}
  \caption{Neutral data server and B2B marketplace architecture for in-vehicle data access in C-ITS supported by ACEA and CLEPA. The green shaded area indicates an in-vehicle data recording, storage and access management implemented as Event Data Recorder for Autonomous Driving (EDR-AD) or Data Storage System for Automatic Car Steering Function (DSSA).}
  \label{fig:acea-proposal}
\end{figure}

Main lobbying and standards groups of automobile industry in European Union -- European Automobile Manufacturers Association (ACEA)\footnote{\href{http://www.acea.be/}{http://www.acea.be/}} and European Association of Automotive Suppliers (CLEPA)\footnote{\href{https://clepa.eu/}{https://clepa.eu/}} have joined forces\footnote{\href{https://www.ereg-association.eu/news-items/acea-and-clepa-join-forces-to-find-a-solution-for-secure-and-safe-access-to-vehicle-data/}{https://www.ereg-association.eu/news-items/acea-and-clepa-join-forces-to-find-a-solution-for-secure-and-safe-access-to-vehicle-data/}} behind the proposal for a Data server / Extended vehicle architecture for allowing access to vehicle data for third-party services and service providers in C-ITS scenarios \citep{acea_position_2016}. The gist is of the proposed architecture (Figure \ref{fig:acea-proposal}) is that data from in-vehicle data recording and storage system (EDR-AD\footnote{Event Data Recorder for Automated Driving (EDR-AD) -- a term used in industry groups GEAR2030 working documents.}/DSSA\footnote{Data Storage Systems of Automatic Car Steering Function (DSSA)-- a term used in UNECE working documents}) will be synchronized with the proprietary vehicle manufacturer's server, which will contain rules for managing authorized services', service providers' and neutral server's access to data. Neutral server will be provided by a third party in a business-to-business (B2B) data marketplace. It will contain a rule engine to send authorized data to specific service providers.

Within the scope of this paper we are mostly interested in the in-vehicle data recording and storage device which will collect and store data of communications between vehicle's ECUs and external messages in C-ITS scenarios. Yet the device cannot be seen as an isolated unit -- it will be an integral part of proposed Extended Vehicle infrastructure (Figure \ref{fig:acea-proposal})

\subsection{Evolving technological landscape and regulatory directions}\label{evolving-directions}

Modern cars are controlled by a large stack of distributed software that executes more than 50 interconnected Electronic Control Units (ECUs) \citep[p. 1]{van_bulck_efficient_2017}. Autonomous vehicles may contain hundreds of such nodes (data source, decision making, actuating, ect.), communications among which will determine the functionality, automation level and behavior of a vehicle and constitute what is called Controller Area Network (CAN). Operation of CAN is safety and mission critical, involves large amount of messages per second and therefore should be designed with security and anti-tampering considerations in mind. While recent standardization efforts address security, no practical solutions are implemented in current cars \citep[p. 1]{van_bulck_efficient_2017}. Considering drastically increasing importance of in-vehicle communications networks in automated driving systems for overall functionality of vehicles and their safety characteristics as well as dependency on out-vehicle communications in C-ITS, there is a clear and urgent need to develop secure architectures of CANs ready for C-ITS scenarios which not only preserve current status quo in automotive security, but largely enhance it.

Results of state-of-the-art research in embedded and distributed systems security and privacy provide embedded protected module architectures with strong security guarantees having application potential for efficient message authentication in vehicular communication networks. For example, Sancus 2.0 low-cost security architecture for networked embedded IoT devices \citep{noorman_sancus_2017} can remotely attest to a software provider that a specific module is running uncompromised and can provide a secure communication channel between software modules and software providers. It supports remote software installation on devices while maintaining strong security guarantees. VulCAN -- an efficient component authentication and software isolation for automotive control networks based on Sancus architecture \citep{van_bulck_efficient_2017} -- brings strong security to automotive computing by featuring message authentication, strong software security and Trusted Computing -- software component isolation and cryptography leveraging hardware-level cryptographic primitives for real-time efficiency of automotive networks. VulCAN architecture solves the problems inherent in current complex vehicle bus systems with many ECUs and gateways to other communication systems that do not offer protection against message injection or replay attacks. Furthermore, it provides efficient and cost-effective implementation of AUTOSAR (major industry standardization partnership\footnote{AUTOSAR (AUTomotive Open System ARchitecture -- \href{https://www.autosar.org}{https://www.autosar.org}) is a worldwide development partnership and industry standardization body of biggest automotive manufacturers, suppliers and other interest groups. It pursues the objective of developing and establishing open standards for automotive electronic control units (ECUs) instrumental for industry-wide efficiency of automotive network design.} compliant CAN authentication protocols for future automotive networks. These technologies are clear and ready candidates for developing CANs for automated vehicles and C-ITS scenarios, notably vehicle data access systems -- see Section \ref{in-vehicle-data-storage-requirements}. 

All communications in C-ITS system could be categorized into \textbf{in-vehicle data/messages} and \textbf{out-vehicle data/messages} by simple criteria of whether a message from an ECU leaves the CAN of a vehicle. In terms of EU Data Privacy and Provenance policy there is a clear separation between the two: 
\begin{enumerate}
  \item Out-vehicle data/messages (CAM or DENM) are considered personal data and therefore fall under regulation of GDPR \citep[p. 4]{article_29_opinion_2017} \citep[p. 27]{c-its_platform_final_2017}, but the concrete legal basis for processing such data under GDPR is not yet determined\footnote{ Four possible legal bases or a combination of them are considered: public interests (art 6(1)e GDPR), performance of a contract (art 6(1)b GDPR), consent (art 6(1)a GDPR) or legitimate interest (art 6(1)f GDPR). Also, as ITS Directive 2010/40/EU allows the European Commission to adopt binding specifications in this field via delegated acts, the mandatory deployment of C-ITS is also considered as an option, but not for the initial deployment in 2019 \citep[p. 5]{article_29_opinion_2017}.}.
  \item Additionally, the ePrivacy Regulation may become relevant for out-vehicle data/messages as the second proposal of the Commission states that all machine to machine communications fall under its scope \citep[p. 6]{article_29_opinion_2017}.
  \item In-vehicle data/messages are exempt from GDPR on the grounds that it is considered a processing of personal data by a natural person in the course of purely personal or household activity\footnote{ Regulation (EU) 2016/679 does not apply to the processing of personal data by a natural person in the course of a purely personal or household activity (art. 2(a) GDPR). This exemption may only be valid when strictly limited to the processing that takes place inside a car, and only if the driver is in full control of the processing within the device. It cannot be valid when the device installed in the cars forwards the data of other nearby cars, be it immediately, or as a result of local processing. In such cases the processing is not limited to a strictly personal activity \citep[p. 9]{article_29_opinion_2017}.}.
\end{enumerate}

There is a clear direction for coming up with a legal framework for including data storage requirements in the type-approval legislation of EU. As per recommendations of High Level Group GEAR 2030, the European Commission should monitor and evaluate the need to revise the Motor Insurance and Product Liability Directives as well as the need for additional EU legal instruments to take into account future development of technologies \cite{dg_grow_gear2030_2017}. The Commission will be developing EU type-approval framework for the certification of automated vehicles, including alternative assessment method and identification of work priorities at the UNECE, EU and Member State levels in 2018. 

The exact requirements for the information that will need to be stored by automated vehicles are in the process of development, yet GEAR 2030 meeting documents include a final draft of Discussion Paper on Event Data Recorders for Automated Driving (EDR-AD) of September 2016\cite{european_association_of_automotive_suppliers_clepa_discussion_2016} which suggests that basically all sensor information as well as communications will have to be recorded for at least limited period of time. It also describes that data should be collected in the following nodes in automated vehicle: data source nodes (e.g. sensors), decision making nodes (e.g. neural network nodes which make decision in automated driving scenarios), actuating system nodes (e.g. braking, throttle) and special data collection node (i.e. "black box"). Data privacy is a key concern that will have to be addressed within the regulatory framework of Data Privacy and Provenance Policy. Furthermore, the provisions for using cryptographic algorithm for security and data tampering preventions are considered. \cite{european_association_of_automotive_suppliers_clepa_discussion_2016} is the basis for data requirements of EDR/AD system considered in the proposal for a solution (section \ref{edr-ad-solution}). 

Since requirements for EDR/AD systems allow certain data to leave an in-vehicle network, it is only reasonable to assume that GDPR and ePrivacy regulations may apply as soon as this happens, e.g. when the right of competent authorities to access the in-vehicle data recorded in EDR/AD systems is exercised.

The specific ongoing standardization projects related to in-vehicle data access systems and, consequently, EDR/AD devices are:
\begin{itemize}
  \item Cyber-security regulations for Intelligent Transportation Systems and Automated Driving (ITS/AD) by UNECE Working Party 29 (Section \ref{governance-bodies}) are projected to be drafted in by the end of first quarter of 2018 and finalized by the middle of 2018. All meeting documents of the specific working group regarding these regulations are available on-line\footnote{\href{https://wiki.unece.org/pages/viewpage.action?pageId=2523344}{https://wiki.unece.org/pages/viewpage.action?pageId=2523344}};
  \item New standartization project by ISO for defining the Extended vehicle concept (Figure \ref{fig:acea-proposal}) involves standards on methodology\footnote{ISO 20077-1:2017 Road Vehicles -- Extended vehicle (ExVe) methodology -- Part 1: General information -- \href{https://www.iso.org/standard/66975.html}{https://www.iso.org/standard/66975.html}}, published in December 2017, 'web services'\footnote{ISO 20077-2:2018 Road Vehicles -- Extended vehicle (ExVe) methodology -- Part 2: Methodology for designing the extended vehicle -- \href{https://www.iso.org/standard/67597.html}{https://www.iso.org/standard/67597.html}}, published in January 2018 and remote diagnostic supportfootnote{ISO/DIS 20080 Road vehicles -- Information for remote diagnostic support -- General requirements, definitions and use cases -- \href{https://www.iso.org/standard/66979.html}{https://www.iso.org/standard/66979.html}}, which is still in "enquiry" phase (i.e. about a year from publishing).
  \item Specific regulations regarding inclusion of data recording and storage systems into type-approval requirements (i.e. UN Regulation 79)\footnote{\href{http://eur-lex.europa.eu/eli/reg/2008/79(2)/oj}{http://eur-lex.europa.eu/eli/reg/2008/79(2)/oj}} are under discussion at UNECE's Committee on Automatically Commanded Steering Function (ACSF)\footnote{\href{https://wiki.unece.org/pages/viewpage.action?pageId=25265606}{https://wiki.unece.org/pages/viewpage.action?pageId=25265606}} and expected to be finalized and published in stages in 2018 and 2019;
  \item Finally, some EU member countries, notably Germany, has passed a law\citep[p. 3]{ranft_freeing_2016}\footnote{\href{http://www.bundestag.de/dokumente/textarchiv/2017/kw13-de-automatisiertes-fahren/499928}{http://www.bundestag.de/dokumente/textarchiv/2017/kw13-de-automatisiertes-fahren/499928}} allowing Level 3 automation level (see Figure \ref{fig:levels_of_automation}) on public roads as long as data recording device to be defined by UNECE is installed\footnote{\href{https://www.reuters.com/article/us-germany-autos-self-driving/germany-adopts-self-driving-vehicles-law-idUSKBN1881HY}{https://www.reuters.com/article/us-germany-autos-self-driving/germany-adopts-self-driving-vehicles-law-idUSKBN1881HY}}. 
\end{itemize} 

This indicates a clear consensus that the Event Data Recorder for Automated Driving will be a regulatory prerequisite for Automated Driving in EU\citep[p. 12]{european_association_of_automotive_suppliers_clepa_discussion_2016} and is on a way to be defined by 2019-2020 for the first implementation of C-ITS in Europe. Starting to develop the EDR/AD devices already now (based on the available documentation, standards and working group documents and considering security, privacy and non-tamperability requirements) is instrumental for successful implementation of the system and a business opportunity. We now turn to describe a proposal for concrete in-vehicle data recording, storage and access management solution and discuss the related specific aspects of regulatory framework and technology availabilities.

\subsection{EDR/AD solution for connected autonomous vehicles}\label{edr-ad-solution}

EDR/AD is a subsystem of a vehicular Controller Area Network which ensures the confidentiality, integrity and availability of data related to operation of a vehicle in order to permit recovery of exact situation following the occurrence of an event (e.g. traffic accident or other incident) or on demand. The discussion about EDR use is still going on in the world, but the need for this technology will soon increase with the introduction of automated vehicles on the road. It is still not clearly defined, how future accidents involving semi- of fully automated vehicles will be investigated and judged, but it is clear that the technology for obtaining evidence about the system's state before or after an accident will be needed. The work at many levels of governance (see Section \ref{evolving-directions}) for standardizing the EDR/AD device is being done in order to clarify which data will be recorded, how it will be accessed and by whom as well as how vehicle and C-ITS will look like to guarantee data consistency.

Exact data recording requirements are still under discussion, yet it is reasonable to assume that autonomous vehicles will be recording considerable part (if not all) of internal and external communications, considering the availability of modern cheap storage technologies. Preliminary discussions on nature of data that will be recorded as well as some projected particularities of EDR/AD architecture at the time of writing can be inferred from working documents only\citep{european_association_of_automotive_suppliers_clepa_discussion_2016}. Nevertheless, such uncertainty does not prevent to develop the system, since it will anyway have to feature  flexible on-demand reconfigurability of data sources, storage volumes and time, as well as their access management and permissions, since the regulatory environment (now and in the future) is too fluid for allowing such rigidity. 

\begin{figure}[H]
  \center
  \noindent\makebox[\linewidth]{\rule{\textwidth}{0.4pt}}
  \includegraphics[width=1\textwidth]{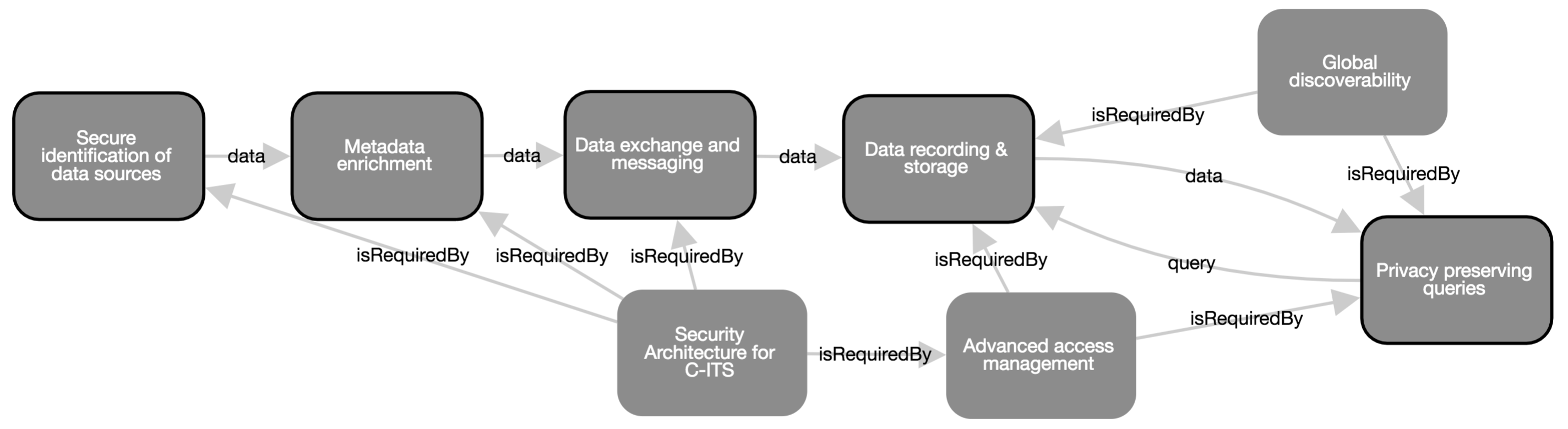}
  \caption{A high level depiction of the components of the EDR/AD for C-ITS solution.}
  \label{fig:edr-ad-solution-components}
\end{figure}

We describe the preliminary architecture of the EDR/AD system in terms of five processes that it has to implement (see Figure \ref{fig:edr-ad-solution-components}).
\begin{description}
  \item[Secure identification of data sources] is needed for ensuring data integrity when the messages originated from different ECUs in an in-vehicular network will be stored and then identified upon retrieval. Security is needed in order to prevent data tampering. Note, that in C-ITS scenarios, data which can influence the behavior of a vehicle also includes CAM messages received from other vehicles and road stations, correct and secure identification of which may have a decisive role for event reconstruction.
  \item[Metadata enrichment] is a process which adds additional semantic data to messages which may be needed for further processing and retrieval. While the exact metadata requirements are not clear at the moment of writing, it will include at least identifier and a timestamp, and also category of a message for access rights management. In case a message leaves an in-vehicular network it may also have to be enriched with the metadata about state of user's consent to share private information (i.e. \textit{consent marker}) \citep[p. 55]{c-its_platform_final_2016}. 
  \item[Data exchange and messaging] is obviously the process that allows the messages to be passed between Electronic Control Units preserving their integrity, security and all metadata. For the solution to be applicable for C-ITS scenarios, data exchange an messaging processes should account for the possibility for some messages to be exchanged with out-vehicle network.
  \item[Data recording \&storage] is the process which takes care of filtering messages that need to be stored according to requirements, and recording them into storage on either dedicated ECU (EDR-AD) or other nodes of the in-vehicle network.
  \item[Access management / privacy preserving querying] is the process which allows for a competent authority or a stakeholder of the ecosystem to access data which it is authorized to access, without  compromising privacy of other stakeholders.
\end{description}

Note, that Extended Vehicle concept provides that data access management rules will be defined and enforced by vehicle manufacturer's or neutral server (Figure \ref{fig:acea-proposal}), therefore the architecture should be taken into account when designing EDR/AD device, which implies the need for cooperation between all stakeholders for designing a it (vehicle manufacturers, component suppliers, cryptography researchers and more). 

The five processes should ensure data security and reliability requirements, including, but not limited to the following (based on \citep[p. 14]{european_association_of_automotive_suppliers_clepa_discussion_2016}):
\begin{itemize}
  \item In terms of \textit{data confidentiality} -- allow to limit the access to authorized persons / institutions only (i.e. limit the access to EDR/AD ECU for the end user and encrypt the stored data). Access management solutions will have to comply to C-ITS trust model and certificate policy based on PKI \citep{c-its_platformm_wg5:_security_&_certification_annex_2016};
  \item With regard to \textit{data integrity} -- protect the system from any modification and tampering (e.g. pair the EDR with the vehicle and signed the stored data). There is a possibility to leverage SANCUS security architecture for these purposes \cite{noorman_sancus_2017,muhlberg_lightweight_2015}.
  \item  In terms of \textit{data availability} -- ensure data recording capability in case of event (e.g. duplicating EDR ECU);
  \item The data securization format could depend on the data type, the storage support and regulatory requirements of EU Data Privacy and Provenance Policy (Section \ref{relevant-legislative-domains});
\end{itemize}

\begin{figure}[H]
  \center
  \noindent\makebox[\linewidth]{\rule{\textwidth}{0.4pt}}
  \includegraphics[width=0.6\textwidth]{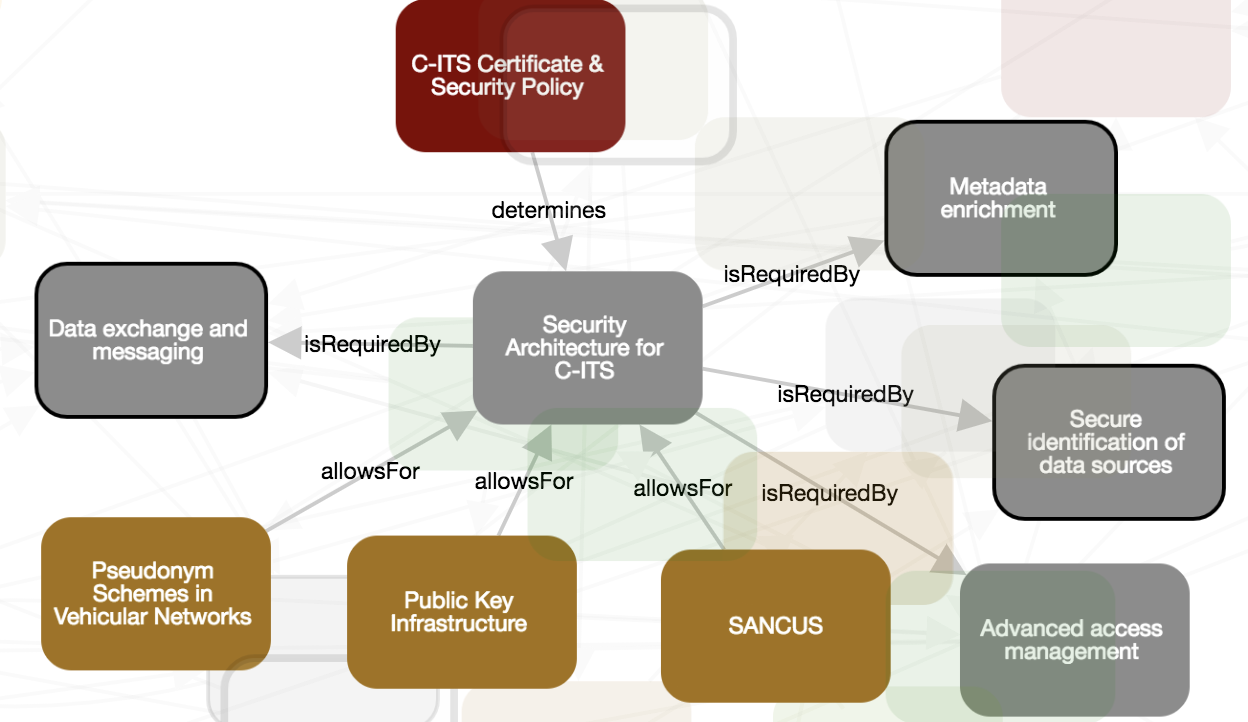}
  \caption{Relations of security architecture for C-ITS and processes of proposed EDR/AD solution.}
  \label{fig:c-its-security-architecture}
\end{figure}

\begin{figure}[h]
  \center
  \noindent\makebox[\linewidth]{\rule{\textwidth}{0.4pt}}
  \includegraphics[width=0.6\textwidth]{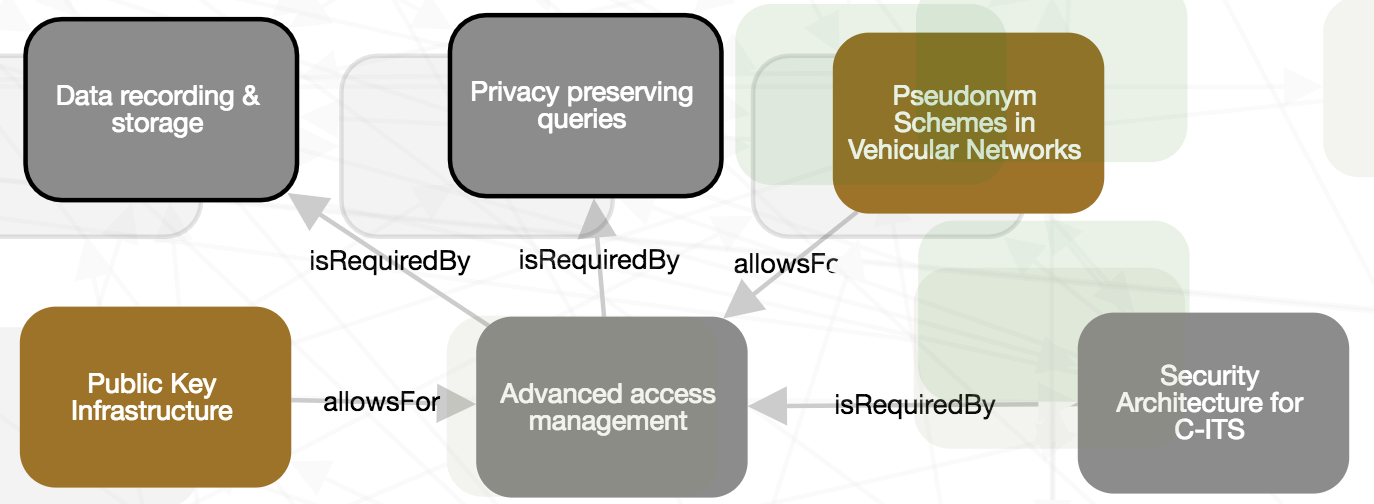}
  \caption{Relations of advanced access management to the processes of proposed EDR/AD solution.}
  \label{fig:advanced-access-management}
\end{figure}

\begin{figure}[h]
  \center
  \noindent\makebox[\linewidth]{\rule{\textwidth}{0.4pt}}
  \includegraphics[width=0.6\textwidth]{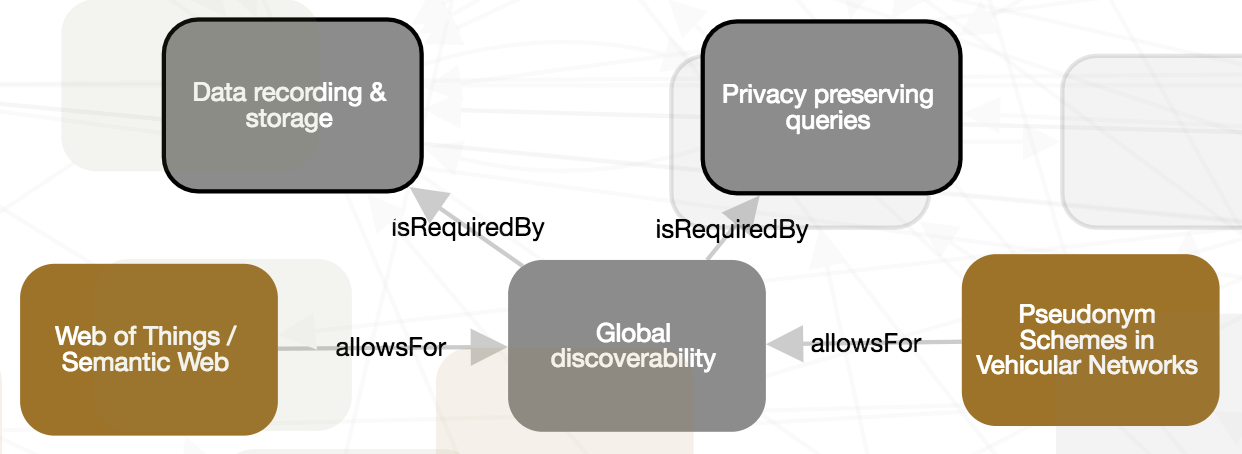}
  \caption{Relations of global discover-ability to processes of proposed EDR/AD solution.}
  \label{fig:global-discoverability}
\end{figure}

From a general perspective, the five processes of EDR/AD solution require to leverage three broad areas:
\begin{inlinelist}
  \item Security architecture for C-ITS, required for data exchange and messaging, meta-data enrichment and secure identification of data sources (Figure \ref{fig:c-its-security-architecture});
  \item Advanced access management solutions, required by data recording \& storage and privacy preserving queries processes (Figure \ref{fig:advanced-access-management})
  \item Global discover-ability solutions, also needed for data recording and storage and privacy preserving querying processes (Figure \ref{fig:global-discoverability}).
\end{inlinelist}

\subsubsection{Security architecture}

Security architecture for the C-ITS is being actively developed by C-ITS Platform and GEAR2030 working groups and is the main pillar of collaborative intelligent transport systems of the future. It is the subject of C-ITS Certificate and Security policy and at the time of writing is being described by two documents:
\begin{inlinelist}
  \item First release of Certificate Policy for Deployment and Operation of European Cooperative Intelligent Transport Systems (C-ITS), June 2017 \cite{c-its_platform_phase_ii_certificate_2017} and
  \item Opinion of Article 29 Working Group on processing personal data in the context of Cooperative Intelligent Transport Systems (C-ITS), October 4, 2017 \cite{article_29_opinion_2017}.
\end{inlinelist}

These documents are mostly concerned with the security and data privacy issues as related to the CAM and DENM messages in V2V (vehicle to vehicle) and V2I (vehicle to infrastructure) communications which are the key aspect of C-ITS functioning. Note, that the policy does not explicitly address security of in-vehicle networks considering it a responsibility of automotive manufacturers. Contrariwise, the proposed solution for EDR/AD for C-ITS integrates aspects of in-vehicle security with the requirement to record and make available driving data as well as C-ITS certificate policy as we believe that such integration is precisely what is needed for the market.

The C-ITS security policy acknowledges that in many scenarios it is important to verify the authenticity and integrity of the messages containing information such as position, velocity and heading of a vehicle, also traffic control messages which are considered for later stages of C-ITS deployment in Europe. The strong authenticity and integrity should allow to assess the trustworthiness of the sent information. The actual identification of the sender of a message may be needed in case of event (e.g. accident) reconstruction and judicial processes, yet the privacy of road users should be maximized at the same time. The developed security architecture is supported by Public Key Infrastructure (PKI) using commonly changing pseudonym certificates \citep[p. 13]{c-its_platform_phase_ii_certificate_2017}. The proposed trust and governance model of PKI includes a number of Enrolment, Authorization and Root Certificate Authorities of each Member State, governed by Trust List Manager and authorized by EU level Certificate Authority. The certificate policy management, PKI authorization management as well as authentication of the Trust List Manager will be a responsibility of Policy Authority -- a role composed by the representatives of public and private stakeholders participating in C-ITS trust model. The policy will be binding for all participants of trusted C-ITS system in Europe. Each vehicle will have to be authenticated within the system at every moment of participation. 

Since C-ITS security architecture will be based on PKI, it is only reasonable to design in-vehicle data access solutions which integrate to the same infrastructure -- i.e. use cryptographic certificates and public/private key pairs for authentication and possibly decryption of encrypted data. Importantly, existing state of the art technologies for ensuring strict security of in-vehicle and out-vehicle communications (e.g. Sancus, VulCAN or similar -- see Section \ref{evolving-directions}) should be leveraged and integrated into the overall C-ITS data flows management architecture for achieving guiding principles set forth by policy makers (Section \ref{in-vehicle-data-storage-requirements}).

\subsubsection{Access management}

An agreement on access to data between all relevant public and private stakeholders should be found, including business model-related question of national or European data licenses, formats, interfaces and addressing privacy concerns \citep[p. 15]{c-its_platform_final_2017}. A technical solution for enabling secure and privacy preserving access to data should be flexible and expendable for accommodating changing requirements of such agreement.

In line with Extended Vehicle concept, access to in-vehicle data will be managed by an external server  -- either vehicle manufacturer's or neutral provider's. Yet, Extended Vehicle will need to provide for at least two more channels of data access -- 
\begin{inlinelist}
  \item peer-to-peer ad-hoc communication with other vehicles, road-side units and traffic managers which is an essential aspect of C-ITS operation;
  \item off-line access via ODB (On-board diagnostics) or ODB II connectors for technical inspection and repair (see Figure \ref{fig:acea-proposal}).
\end{inlinelist}
In order to keep data integrity, all three access points to in-vehicle data -- external server, ad hoc communication and ODB ports -- will have to by integrated via EDB/AD subsystem of vehicular CAN. This is especially relevant to ad-hoc communications as it is the basis for C-ITS. Therefore, Extended Vehicle concept and architecture does not cover all in-vehicle data access issues. Most importantly, it does not provide solutions to the data security and authenticity aspects of C-ITS.

\subsubsection{Global discover-ability}

The requirement of global discover-ability is the least obvious in the proposed solution and may not be needed for the use cases of the first deployment of C-ITS. Yet it is still included into the solution because it should be explicitly addressed during the design and may be needed for the use cases beyond first deployment (e.g. using cameras and sensors of different vehicles or road-side units). Global discover-ability implies two functionalities:
\begin{inlinelist}
  \item as soon as the data leaves an in-vehicle network, it will have to be uniquely identified and related to a vehicle (via PKI or pseudonym certificates) in order to ensure certain aspects of Data Privacy and Provenance policy (e.g. right to be forgotten principle);
  \item in C-ITS use cases where a vehicle will rely on information received from another vehicles or roadside equipment or even will use their sensors remotely, a system of devices broadcasting the information about their capabilities and availability to share information.
\end{inlinelist}

\section{Conclusion}\label{conclusion}

The transport sector is in the process of being rapidly and fundamentally reshaped by new emerging technologies, including autonomous driving and collective intelligent transportation systems. This reshaping promises huge economic and social benefits yet brings equally huge challenges in terms of secure and safe technology development, its smooth integration to social fabric and the adaptation of legal and regulatory framework. In the previous article we have developed Policy Scan and Technology Strategy Design methodology \citep{veitas_policy_2018} for identification of concrete societal expectations and problems and technological availabilities that can mitigate them in the domain of autonomous driving and smart mobility. As a result we have identified technology having a clear place in emerging technological and regulatory landscape as well as market deployment potential -- Event Data Recorder for Autonomous Driving (EDR/AD).

In this article, particularities of the context of EDR/AD related technology availabilities, regulatory initiatives and problematics of the large scale industry transition at European and international level are presented and discussed in depth. EDR/AD is one of the technological solutions which, if designed by integrating knowledge existing in the industry and academia, can help to design a robust and secure in-vehicle data recording, storage and access management solutions which will be a regulatory prerequisite for the deployment of autonomous vehicles.

Transformation of transport sector involves many stakeholders and interest groups and it is important to understand that no one of them has a decisive power to shape the direction of the transformation (albeit some of them have more power than others). Guided by Policy Scan and Technology Strategy design methodology \citep{veitas_policy_2018} we explicate ecosystem of actors (industry participants, policy makers, technology availabilities and more) around EDR/AD solution and its place in the C-ITS. The aim of this explication is to pave ground for collaboration of relevant stakeholders for developing and implementing concrete technology of EDR/AD, considering its short and long term effects on the transformation of smart mobility systems, regulatory aspects and business opportunities.

Further steps for developing EDR/AD as envisioned in this paper involves building collaboration between vehicle manufacturers and security researchers in order to bring state of the art security technologies to the industry, proposing concrete use cases for prototyping and security analysis and developing a proof of concept demonstrator of technology that is market-deployment ready. Considering the fast pace of automated driving technology development, plans to deploy collaborative intelligent transportation systems in the short term and serious shortcomings in security and non-tamperability of current vehicular networks, we conclude that development of EDR/AD is an urgent need for industry.


\printbibliography[heading=bibintoc]
\end{document}